\documentclass[prl,twocolumn]{revtex4}

\pdfoutput=1

\usepackage{amsmath} 

\usepackage{graphicx} 
\usepackage{setspace}

\usepackage{enumitem}
\usepackage{booktabs}

\begin{document}

\title{Microscope Project for Undergraduate Laboratories}
\author{Rachel Kemp}
\affiliation{University of Massachusetts Amherst, Department of Biochemistry and Molecular Biology, Amherst,
MA 01003}
\author{Alexander Chippendale}
\affiliation{University of Massachusetts Amherst, Department of Physics, Amherst,
MA 01003}
\author{Monica Harrelson}
\affiliation{University of Massachusetts Amherst, Department of Physics, Amherst,
MA 01003}
\author{Jennifer Shumway}
\affiliation{University of Massachusetts Amherst, Department of Physics, Amherst,
MA 01003}
\author{Amanda Tan}
\affiliation{University of Massachusetts Amherst, Department of Physics, Amherst,
MA 01003}
\author{Sarah Zuraw}
\affiliation{University of Massachusetts Amherst, Department of Physics, Amherst,
MA 01003}
\author{Jennifer L. Ross}
\affiliation{University of Massachusetts Amherst, Department of Physics, Amherst,
MA 01003}
\email{rossj@physics.umass.edu} 
\date{\today}

\begin{abstract}
Optics is an important subfield of physics required for instrument design and used in a variety of other disciplines, including materials science, physics, and life sciences such as developmental biology and cell biology. It is important to educate students from a variety of disciplines and backgrounds in the basics of optics in order to train the next generation of interdisciplinary researchers and instrumentalists who will push the boundaries of discovery. In this paper, we present an experimental system developed to teach students in the basics of geometric optics, including ray and wave optics. The students learn these concepts through designing, building, and testing a home-built light microscope made from component parts. We describe the experimental equipment and basic measurements students can perform to learn principles, technique, accuracy, and resolution of measurement.  Students find the magnification and test the resolution of the microscope system they build. The system is open and versatile to allow advanced building projects, such as epi-fluorescence, total internal reflection fluorescence, and optical trapping. We have used this equipment in an optics course, an advanced laboratory course, and graduate-level training modules. 

\textbf{Author's note:} This manuscript was reviewed and accepted by the scientific reviewers and editors at American Journal of Physics. Unfortunately, I was informed that I needed to include my course materials in the supplement, which could negate my copyright. Further, I refused to make further non-scientific editorial edits. It was thus in the 2\% of papers that are scientifically accepted and editorially rejected. This occurred after two years, which I also have objection to, since the point was to create a document to enable others to offer a class that my students and I find valuable. I apologize to the scientific and educational community for these delays. I have decided to make it freely available here on arXiv without further delay.
 \end{abstract}

\maketitle

\section{Introduction}
 The microscope is a familiar piece of equipment to many disciplines, including physics, materials science, chemistry, and the life sciences. The microscope uses the fundamental aspects of basic ray and wave optics that are important for everyday measurement in these fields, making microscopes ideal optical system to offer opportunities to educate students in the basic concepts of optics and measurement principles, techniques, accuracy, and resolution. For instance, optical systems, the resolution is easily and directly measured and can be compared to theoretical predictions. \cite{Pedrotti} Such experiments can offer a number of meaningful hands-on activities of measurement accuracy. 
 
 Despite its commonplace appearance, modern microscopes can be quite complicated and modern microscopy techniques can be sophisticated, using advances optical concepts to add novel functionalities to the basic system. Even concepts such as Fourier optics, holographic optics, and quantum optical systems, such as pulsed lasers, are used in high-end, modern microscopes.  Further, recent, advanced analysis techniques are challenging ``resolution limits,''\cite{Huang} and are being implemented by a number of scientists to perform novel experiments. In order to fully understand the abilities and quantitative nature of their microscope, and offer opportunities to improve optical design of such systems, students need to be educated in basic optics. Here, we describe an experimental system and measurements students can use in order to learn basic optics design and construction to build a working light microscope. Using their home-built system, students can probe the meaning of measurement accuracy and resolution.  

Research-based education techniques inform us that training students using hands-on, kinesthetic activities is the best way to make true learning gains in a subject.\cite{Prince} Further, giving students the ability to work on a long-term project (project-based learning) to build an microscope that they then use is powerful not only for learning, but also for stimulating student interest in the subject.\cite{Strobel} Thus, we sought to create a laboratory apparatus that would allow both hands-on learning on a semester-long project combined with modern and advanced equipment such as is used in biophysical research labs that build novel microscopes.

 \subsection{Teaching Venues}
We have used the basic system we describe here in three teaching venues including (1) a semester-long optics course originally for physics majors that was cross-listed with a biology course to enable life science, physical science, and engineering students to take the course simultaneously; (2) a semester-long advanced laboratory course for physics majors where students perform four experiments in groups, give presentations, and write manuscripts with the focus on measurement techniques and experimental design; and (3) a two-day laboratory module on optics for graduate students from a variety of disciplines including life science, physical science, and engineering. In all three venues, the basics of geometric optics were taught through students actively building and testing the light microscope, although the apparatus was used slightly differently and there are scaffolded exercises for each (available upon request). 

\subsubsection{Undergraduate Optics Course}
For the semester-long optics course, the course had 10 - 14 students and four complete experimental systems with three to four students per group working on a single microscope system. The class was held in a room with black-top benches and no optical benches. We found that having three students per experimental system was optimal is such a room arrangement. The groups of four did not have enough room around the equipment for all four students to access the system simultaneously. During the class, these 12 students were assisted by one professor and one teaching assistant who strolled through the room helping people. 

For this course, the project of building a light microscope was broken up into smaller modules with significant scaffolding. For each module, students worked in small groups to design and test the components described below.  Below, we describe the individual modules of the microscope including the condenser (Section II. B.), imaging path with a camera (Section II. C.). We also describe the measurements the students need to make with their completed microscope including calibrating the magnification and testing the theoretical resolution limits of their system (Section III). In this course, the students spent the last two to three weeks working on an additional "advanced" optical system or project of their choosing (Sections IV. A. and B.). We describe two examples where students built an epi-illumination path for fluorescence or used the concepts of a priori knowledge of the structure of the sample to increase the accuracy of measurement similar to the basic techniques used in ``super-resolution'' microscopy. 

At the conclusion of each section (building the condenser, completing the microscope, building an advanced system), students presented their work in poster format to each other and to faculty and graduate students from a variety of departments. They were required to discuss certain aspects of their design and measurements including ray diagrams and matrix methods calculations of their designs. We are happy to provide more information, specific examples, homework assignments, or scaffolding (worksheets) used to guide students upon request by emailing the corresponding author.
 
\subsubsection{Undergraduate Advanced Laboratory Course for Physics Majors}
In the semester-long advanced laboratory course for junior and senior physics majors, the apparatus was used for one of the four required experiments students had to perform in the course. Groups consisted of two to three students, and two experimental systems were used simultaneously to allow up to 6 students to work on microscope building and measurement at the same time. For this course, there is a maximum of 12 students for one professor and one TA. 

Students used published articles as guides to building the equipment. There is no scaffolding for this course, because one of the goals of the course is to have student engage literature and use it for experimental design and measurement. For this course, which focuses on measurement technique, students made measurements of the magnification and resolution of their system over two to three days. They presented their work in a manuscript and a 20-minute presentation. Many of the results presented in this article are from their data working on the microscope.

\subsubsection{Graduate Laboratory Modules}
We used the same concepts and experimental equipment for several two-day laboratory modules for advanced students. The systems have been used by graduate student on campus as well as exported it to the Analytical and Quantitative Light Microscope (AQLM) course offered at the Marine Biological Laboratory in Woods Hole, MA. Because we are teaching all of basic optics in a short time span, the scaffolding for this module is extensive (available upon request), and lectures were kept to a minimum (three 10 - 30 minute lectures mostly to motivate the hands-on activities).  

At AQLM, we had 34 students working on 8 experimental systems simultaneously in a single, large laboratory room. The systems were stationed on low blacktop countertops with chairs around them. Some of the groups had 5 students with one apparatus, which is on the high side, but students were mature and took turns working with the equipment. Each apparatus had one to two industrial teaching faculty working with each team one-on-one. Students learned concepts important in the optics of microscopes such as collimation, infinity space, focusing, image formation, measuring magnification, and resolution.

 \subsection{Scope of the Project}
In this paper, we present the experimental system to train students in the basics of ray and wave optics using the light microscope as a unifying project. We present the components, assembled microscope equipment, and basic measurements that students have used to learn geometric optics.  Students design, build, and test their own home-built optical microscope. We demonstrate how students can find the magnification and test the resolution of the system they build. Through these experiments, students learn the concepts of accuracy and precision in measurement techniques and how to estimate their uncertainty. 

\section{Experimental Equipment}
 In this section, we detail how the students have designed and built transmitted light microscopes using component parts. 
 
 \subsection{Components}
 The optical components required to build a working, modern light microscope are outlined in Table ~\ref{Components} at the end of this document. All components were purchased from ThorLabs, and part numbers and images are given to serve as examples and not to specifically endorse a particular vendor. Because the parts are durable and basic, recycled or even 3D printed parts could be used.\cite{Dvorak, Zhang} In this section, we justify some of the choices we made for components. The optical breadboard with 1/4"-20 tapped holes in a 1-inch square grid was used as a platform to build the microscope if an optical table is unavailable. Using small breadboards can save space. One dimension needs to be at least 2 feet long so that the optical train is long enough to achieve the correct magnification, but the other dimension can be as thin as 6 inches for a standard transmitted light microscope. Rubber feet are used under the breadboard to elevate and stabilize the system against vibration. Using breadboards allows the systems to be packed away during alternate semesters or transported for other labs to use. 
 
 To create the optical train, we chose to use dove-tail rails that were half an inch wide. Rails are common in other pre-fabricated optical laboratory systems for student labs, such as those that can be purchased from PASCO, because they allow easy positioning of the optical components along the same optical axis for students who are not adept at aligning. Unlike large, bulky systems from PASCO, the half-inch rails are inexpensive and are authentically similar to what one might find in a research laboratory. Further, other rail systems, including inexpensive home-built systems,\cite{Dvorak} or those printed with new 3D printers would also work.\cite{Zhang} 
 
 We chose rails over cages for confining the optical train because cages are more difficult to add and remove new components. Since the students do a lot of trial and error, the cage system becomes hindering to ad hoc additions or subtractions of components. We also chose rails over affixing the components to the optical board with kinematic mounts because new students often have difficulty aligning freely mobile optical components. The rail is more rigid and easier to align for novice users. Thus, although there are a large number of options for optical systems, the rail system was preferred. Optical components were mounted to the rail on half-inch posts with post-holders attached to dove-tail rail cars. Short posts, two inches long, are preferred for stability, but we used 3 inch and 4 inch posts without detrimental effects on the stability of the system. The ability to use what is available without negative effects makes this laboratory easily duplicatable and adaptable.
 
 The light source was a white light emitting diode (LED) with a variable power supply. The variable power supply is a simple potentiometer that allows the intensity to be changed. The LED does not come with a collecting lens, because we want the students to use it as a point source without a lens. A collecting lens is a crucial component of the condenser they need to design and build, so purchasing one with the lens attached would defeat the point of part of the exercise.  It serves well as a source for mini-laboratories on image formation as well as the lamp for the transmitted light path on the microscope. Although we do not use other filters such as neutral density filters, a monochromatic interference filter, or a diffuser, those components are often used in a condenser and can be added to alter the construction of the condenser (described below).  
 
 Plano-convex and achromatic lenses of various focal lengths were purchased from ThorLabs or Edmund Optics to allow students to tinker with a different lenses. Extras lenses with focal lengths preferred by the students, such as 30 mm, 50 mm, and 100 mm, were also purchased to ensure there were enough. All lenses were one-inch in diameter held in one-inch lens holders mounted to half-inch diameter posts, as described above.  
 
 Objectives of various magnifications were donated from prior laboratory courses on microscopy for use in this course. Both infinity-corrected and 160 mm focal length objectives were used, as described below. A special adapter was purchased to mount the objective with RMS threads to the lens holders. The sample holder was a simple pressure clip, although a variety of other holders can be purchased or made. Some small translation stages with half inch to one inch range were used for fine focus motion. The translation stages can be placed under the objective or the sample holder on the rail. 
 
 Modern light microscopes use CCD or CMOS cameras to capture images (Fig.~\ref{fig:modernmicroscope}). We chose inexpensive but small and powerful CMOS cameras to capture images with a USB link to a student's laptop. The cameras are simple bare-chips of pixels. We do not recommend that the students use other cameras such as the camera from their cellular phone, for instance, because those have lenses built in to allow imaging of objects at a distance (see below). 
 
 The microscope designs we describe here do not allow the students to directly observe the image through the eyepiece, but only allows imaging onto the camera. This is mainly because we are building the system horizontal on the bench and leaning over to put a face in the beam path is unsafe. Here we do not present any systems with lasers, but the system we develop can easily be modified to an optical tweezer or total internal reflection fluorescence microscope where lasers are used. When lasers are implemented laser safety guidelines must be followed, including keeping your eyes out of the beam path, which should stay horizontal to the bread board.

An instructor could have students use their camera phones to capture the images, although we have not. It should be stressed to the students that such cameras already have a lens in front of the detector. This unknown lens before the camera detector takes light that is collimated and focuses it onto the detector. For educational purposes, the phone's camera system can be likened to the eye that also has a built in lens (cornea) in front of the detector (retina). In order to use a phone camera or the eye as a detector, the students must add optics to collimate the image light into the camera. This is similar to adding an eyepiece to the microscope. The eyepiece that is added can alter the magnification of the system and should be included in the discussions with students about calibration, magnification, and resolution (below).  
 
\subsection{Condenser}
\begin{figure}[h!]
\centering
\includegraphics[width=3.5 in]{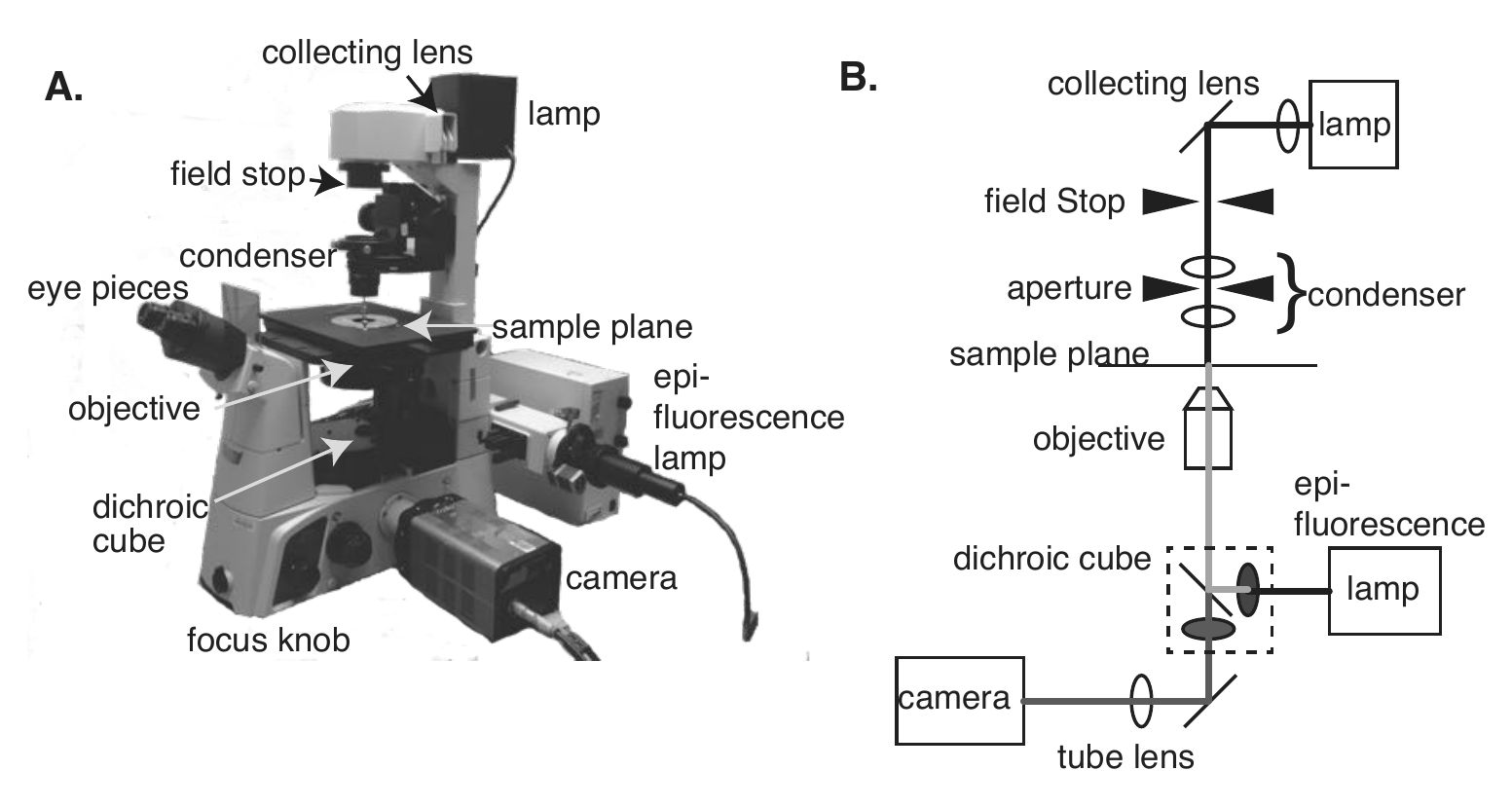}
\caption{Modern microscopes for fluorescence imaging of single molecules, cells, and tissues are of the inverted style with the objective under the sample stage. \textbf{A.} We show an photograph of a modern microscope with the parts of the microscope highlighted that are the same as those the students will build. \textbf{B.} Schematic diagram of the modern microscope with the components drawn out as models to make the paths for illumination and imaging clearer.}
\label{fig:modernmicroscope}
\end{figure}

The microscope condenser is an optical system that controls the brightness and the location of the illumination on the sample and can ultimately influence the resolution of the image obtained. Control is created using irises as apertures and field stops. Unlike simple microscopes that use ambient light directed onto the sample, in a modern, inverted microscope, the condenser system is broken up into a number of pieces between the lamp and the sample. A collecting lens and the field stop are attached in the housing near the lamp, and the aperture is attached to the lenses of the condenser that can be positioned on a rail mounted vertically on the arm of the microscope (Fig.~\ref{fig:modernmicroscope}). 

\begin{figure}[h!]
\centering
\includegraphics[width=3.5 in]{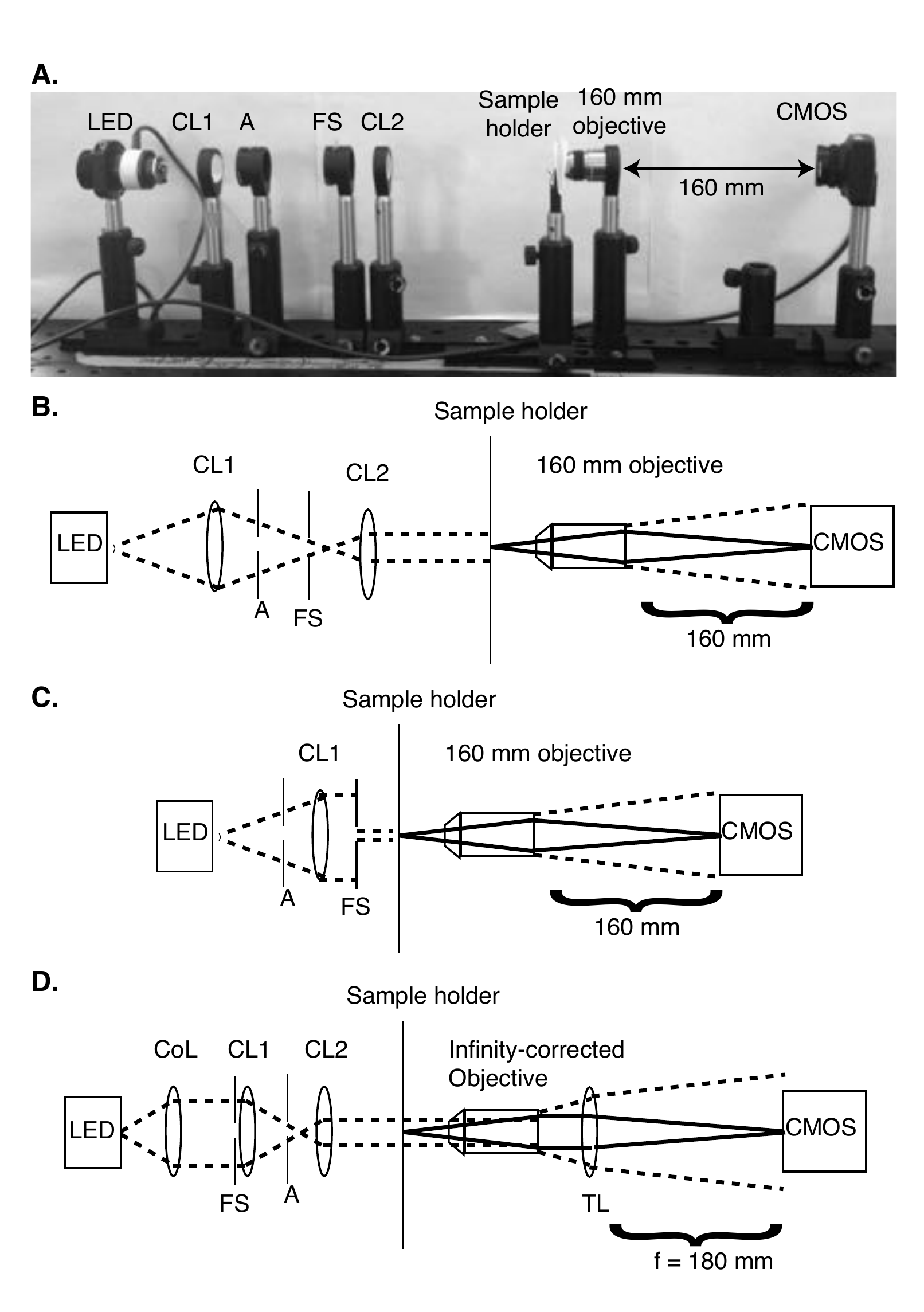}
\caption{Examples of student-designed optical trains for transmitted light microscope. \textbf{A.} Photograph of a working transmitted light microscope created by the students. The left side shows a light emitting diode (LED) as the light source, an iris to act as an aperture (A), two lenses act as a condenser lens (CL1, CL2) to collect and collimate the light at the sample, a second iris to act as a field stop (FS), and then the sample holder. After the sample, a 20x, 160 mm objective is used to create an image onto the CMOS camera 160 mm away from the back of the objective.  \textbf{B.} A schematic diagram of the optical train of the microscope constructed and shown in part \textbf{A.}.\textbf{C.} A schematic diagram of a different design for a microscope condenser where the condenser consists of one lens, and the aperture and field stop irises are in alternate locations.  \textbf{D.} A schematic diagram of a different design for a microscope condenser and imaging path. In this design, the condenser consists of three lenses, and the aperture and field stop irises are in alternate locations. The imaging path between the objective and the camera is also altered because the students used an infinity-corrected objective that requires a tube lens (TL) with a focal length of 180 mm. }
\label{fig:opticalpath}
\end{figure}

In the student designed microscope, there is also a collecting lens and an iris that acts as a field stop. The position of the field stop is variable and determined empirically by the students (Fig.~\ref{fig:opticalpath}). Through their trial and error, they discover the best placement of the iris and link the empirical location with the theoretical discussions of field stops from the textbook or available lecture material. The students find that placing the iris at a location where the rays are collimated often works easiest for creating a field stop. Other correct placements are any location that is a conjugate plane to the sample plane. Such locations will allow the imaging of the field stop onto the sample plane because of their conjugate status. 

A second iris that acts as an aperture. An aperture is an iris that cuts the angle of rays that are continuing through the system. The condenser aperture controls the numerical aperture of the condenser system and ultimately the resolution of the image.  The numerical aperture is defined as:

\begin{equation}
NA = n \sin(\theta_{max}).
\end{equation}
where NA is the numerical aperture, $n$ is the index of refraction of the glass, and $\theta_{max}$ is the maximum angle of rays that can propagate through the optical system. Because the iris that acts as the aperture is variable, the students can adjust the numerical aperture of the condenser (see resolution below). Students of the optics course are required to calculate the numerical aperture of their condenser for the largest and smallest iris radii.

Again, in order to determine the optimal location for the aperture, the students tinker with the position that will reduce the angle of the rays  (Fig.~\ref{fig:opticalpath}). The aperture iris is best placed in a location where the rays from the LED light source converge or diverge (near a conjugate image plane to the LED). Since the LED is not imaged at the sample, the rays from the image of the LED are convolved at the sample plane. Some locations can cut the rays, but are not effective unless the iris is set to a small radius. Although the iris is acting as an aperture here, the placement is not optimal to give maximum control of the numerical aperture using the variable iris. In other words, there are many settings for the variable iris where another aperture (such as the edge of a lens holder) is acting as the aperture. Students are encouraged to re-position the aperture until the iris controls the intensity of light at any radius. 

Overall, when the light encounters the sample, the best illumination is even across the sample. It is stressed to the students that they should not be making an image of the LED lamp on the sample, or else they will create a real image of the LED onto the camera ultimately. By building the microscope in this way, students are automatically creating an objective that is aligned in K{\"o}hler illumination. On a modern inverted microscope, where the condenser lenses can move up and down, the condenser location must be set to K{\"o}hler illumination (Fig.~\ref{fig:modernmicroscope}), but the homebuilt microscope is built in K{\"o}hler illumination and not adjusted once set. 

In order to design a good condenser, the students must understand and use the concepts of: image formation, focal planes, collimation, apertures and field stops. The project of designing the condenser can be as simple or as complex as you need for your class.  In the optics course, students are required to create exact ray diagrams and perform analytical ray tracing with matrix methods of the condensers they designed. They must demonstrate that their condensers can control the light location with the field stop and the intensity with the aperture at the sample plane. As described above, a good condenser does not create an image of the LED at the sample plane. The condenser does make an image of the field stop iris in K{\"o}hler alignment, allowing the students to assess if the field stop was correctly placed. Students from the life science field said that applying their knowledge of K{\"o}hler alignment to create the condenser is one of the most challenging parts, but also the most exciting when they find a configuration that works.

The condenser designs of students are the most variable and interesting part of the microscope design. Students in prior years have used one, two, or three lenses to create evenly illuminated light on the sample plane (Fig.~\ref{fig:opticalpath}). All designs have two irises - one for a field stop and one for an aperture (Fig.~\ref{fig:opticalpath}). Any of these designs are ``correct" and functional as long as they  control the field of illumination, the brightness of illumination, and do not create an image of the LED on the sample plane.  This is a beautiful part of optics - if it works, it is correct, and there are many ways to solve the same problems. 

We demonstrate three example optical diagrams with different condensers with one, two, and three lenses (Fig.~\ref{fig:opticalpath}, B, C). We also show a photograph of an actual microscope that was built by students in the optics course where they used two lenses for the condenser (Fig.~\ref{fig:opticalpath}, A). 

\subsection{Imaging Path}
After creating the condenser, students next have to use an objective to create an image onto the CMOS camera chip. One concept that often stumps students when designing the imaging path is that the LED is not the object we are interested in imaging. Instead, we are interested in making an image of the sample that we place in the sample plane. This can be confusing because the students must use the illumination light of the LED, but we do not want to image the LED itself.

In order to build the imaging path, it is helpful if the students have a sample that they are trying to image. We found that it is easiest if the students use a slide they have written on with permanent ink as a sample to start. This slide is mounted in a fixed, pressure-mount (Table ~\ref{Components}).  The slide is placed in the sample plane of the condenser exactly where the field stops and apertures are aligned to control the location and brightness of the light, in K{\"o}hler illumination.  

For the optics course, where the students have a long time to work, they are encouraged to try to create images of the sample slide using first a single plano-convex lens that creates an image into a sheet of paper. When the paper is replaced by the CMOS camera, the image can be observed on the computer. Students can also use two plano-convex lenses to make an image into the CMOS camera. The two-lens system, such as a telescope, helps the students understand a multi-lens system, much like the objective. 

Finally, students use an objective with relatively low magnification (4x or 5x) and 160 mm focal length to create an image onto the paper and ultimately onto the CMOS camera detector. The objective is mounted into a holder using an adapter ring to adapt the RMS threads to the threads of the lens holders (Table ~\ref{Components}). Movement of the position of the objective along the rail works for gross focus. A small, one-dimensional translation stage under the objective holder allows for fine focus (Table ~\ref{Components}). With the 160 mm objective in place, students can change the sample from the slide with the permanent marker to a micrometer scale. They can use the micrometer scale to check the magnification and resolution (see Experimental Results). 

Depending on the purpose of the project for the course, we also encourage students to replace the 160 mm objective with an infinity-corrected objective. Infinity-corrected objectives are used on modern microscopes because they have an infinity space of collimated rays at the back of the objective between the objective and the tube lens. This region is where optical components that help generate contrast can be placed. For instance, an iris placed here can alter the numerical aperture size of the objective. Because the image is formed at ``infinity'' they are often able to make an image after thc condenser that is very far away from the objective. 

Practically speaking, infinity corrected objectives require a tube lens of specific focal length to achieve the correct resolution and magnification. Each manufacturer's design requires a specific tube lens (Fig.~\ref{fig:modernmicroscope}).  We encourage students to determine how infinity-corrected lenses work through trial and error. They often are missing the tube lens and have trouble creating an image or choose an incorrect tube lens and obtain the wrong magnification (see Experimental Results, below). If they are missing the tube lens, they can usually still create an image far from the objective, effectively at infinity.

Designing and building the imaging path reaffirms the concepts of image formation and ray tracing, while introducing new concepts such as magnification, diffraction, and resolution. Below, we describe how the students can test these concepts with their microscope. Because we are imaging with a camera, the image formation is very straight forward. For instance, when students create an image a piece of paper and then move the camera detector to that position, it reinforces the concept of real image formation. For students who are familiar with microscopes where they use an eyepiece, it is stressed that the optical train to the camera is different than that going to the eyepieces and ultimately to the eye. In particular, the eyepieces on a commercial microscope act to collimate the light again, because the eye has a built in lens in front of its detector, the retina.

\section{Experimental Results}

\subsection{Measuring Magnification}
Once the students are able to create an image onto the CMOS, focus it, and take a picture with the camera software, they need to determine the magnification of their optical system and compare it with the magnification listed on the objective. Individual objectives are made with specific microscope optical systems in mind. In particular, each microscope manufacturer has slightly different optical requirements for the objective including the distance from the objective to the sample (working distance), distance from the back of the objective to the camera (focal distance), need for a ``tube lens'' (for infinity-corrected objectives), and the focal length of that tube lens. All of these differences can alter the magnification achieved by a objective, and thus the students need to carefully calibrate the microscope they build. Ultimately, the objectives are built for specific magnifications in order to achieve a certain numerical aperture and resolution (see below), and thus the students should strive to create the correct magnification for the objective they have. 

Magnification calibration can be achieved by imaging a micrometer, graticule, or diffraction grating with a known spacing between the rulings. Once the image is in focus and saved, we suggest the student open the image using ImageJ. ImageJ is a free Java-based software that works on both PC and Mac platforms. The students can open the image, regardless of the format of the image, adjust the brightness and contrast, and zoom in or zoom out on the image, crop, and save in different file formats. In order to measure the distance between two lines, most students use the Measure tool in the Analyze menu. The types of measurements that are recorded can be set in the Set Measurements menus of the Analyze menu, so that the length can be reported in ImageJ and saved as a text file. The students need to know the units of their measurements, which will usually be pixels. They can draw a line region of interest (ROI) between two lines and measure the distance in pixels between these two lines. Using the known distance between the two lines and the measured number of pixels, they can find the size of a single pixel in nm.

Students will use the micrometer to measure the size of a pixel, but they often become confused because they do not know how to equate the pixel size to a magnification. In order to determine if the size of the pixel they measure is correct, they must know the actual physical size of a pixel. Camera pixels are given in the specifications for the camera, which can be found in the documentation for your camera. For the CMOS cameras we use, the raw pixel size is 5.2 $\mu$m x 5.2 $\mu$m, square. We do not tell the student the pixel size, but insist that they search the documentation for the camera to find it themselves. It is an important experimental skill to know how to find the information for your own equipment, whether they find it online or from the paper documentation that comes with the equipment.

Once the students know the size of the pixel, they can compare the known pixel size to the measured pixel size. Since the image is being magnified, and the physical pixel size is staying the same, we expect the new pixel size to be smaller by a factor equal to the magnification. The magnification can be calculated as:
\begin{equation}
M = \frac{P_{known}}{P_{measured}}.
\end{equation}
where M is the magnification, $P_{known}$ is the known pixel size of the camera (5.2 $\mu$m), and $P_{measured}$ is the measured pixel size from ImageJ in $\mu$m. 

\begin{figure}[h!]
\centering
\includegraphics[width=3.5 in]{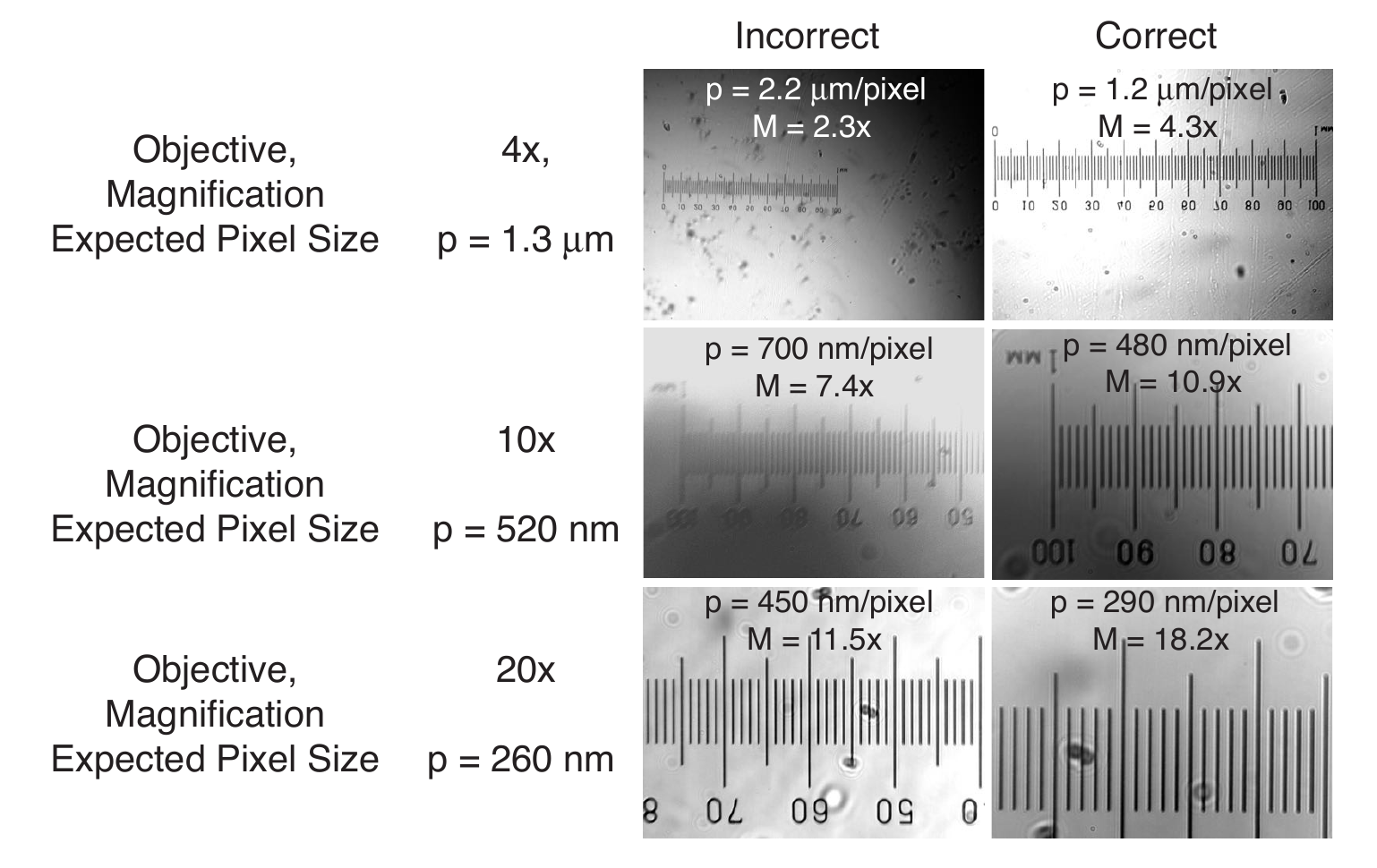}
\caption{Example of correct and incorrect imaging of a 1 mm scale bar with 100, 10 $\mu$m markings using three different objectives. Incorrect placement of the objective or incorrect tube lens focal lengths will result in incorrect magnifications. For the 4x, 160 mm objective, the correct magnification should be 4x and the correct pixel size after magnification of a 5.2 $\mu$m pixel is 1.3 $\mu$m. Incorrect placement of the CMOS camera from the back of the objective results in an incorrect pixel size of 2.2 $\mu$m and an incorrect magnification of 2.3x. 
For the 10x objective, the correct pixel size is 520 nm. The incorrect setting for the 10x is due to incorrect camera placement with respect to the 160 mm objective to give a pixel size of 700 nm and a magnification of 7.4x. 
The 20x objective is an infinity-corrected objective, and the correct pixel size is 260 nm. In the incorrect image a tube lens with focal length of 100 mm is used instead of the correct 180 mm tube lens to give a pixel size of 450 nm and a magnification of 11.5x. Although the correct focal length tube lens is 180 mm, we did not have the correct lens, so a 175 mm lens was used. This lens achieves a pixel size of 290 nm and a magnification of 18.2x.}
\label{fig:magnification}
\end{figure}

Students often find that the magnification is not what they expect given the power of the objective they have chosen (Fig.~\ref{fig:magnification}). There are several reasons why this can occur.  First, if they use a 160 mm focal length objective, they often do not put the CMOS chip at the required 160 mm from the objective on their first try. If the CMOS is at a different distance from the detector, the students will still be able to focus the image, but the magnification will be incorrect. Even a small deviation in the placement of the CMOS detector will result in differences in the magnification (Fig.~\ref{fig:magnification}).

Second, if students use an infinity-corrected objective, the magnification will depend on the specific tube lens that is forming the image onto the camera (Fig.~\ref{fig:magnification}). The tube lens length is based on the manufacturer. We suggest that the students go online and learn about tube lenses and determine the needed tube lens length.  Students will measure the wrong magnification because they often choose a random lens. They will still be able to focus the image and make a measurement, but the magnification will be incorrect (Fig.~\ref{fig:magnification}). We find that the trial and error of measuring and failure allows them to tinker further to correct their mistakes. These missteps and corrections often occur in a real research setting, thus replicating a more realistic research environment. Further, mistakes promote tinkering and problem-solving strategies.\cite{Prince} 

\subsection{Measuring Resolution}
\begin{figure}[h!]
\centering
\includegraphics[width=3.5 in]{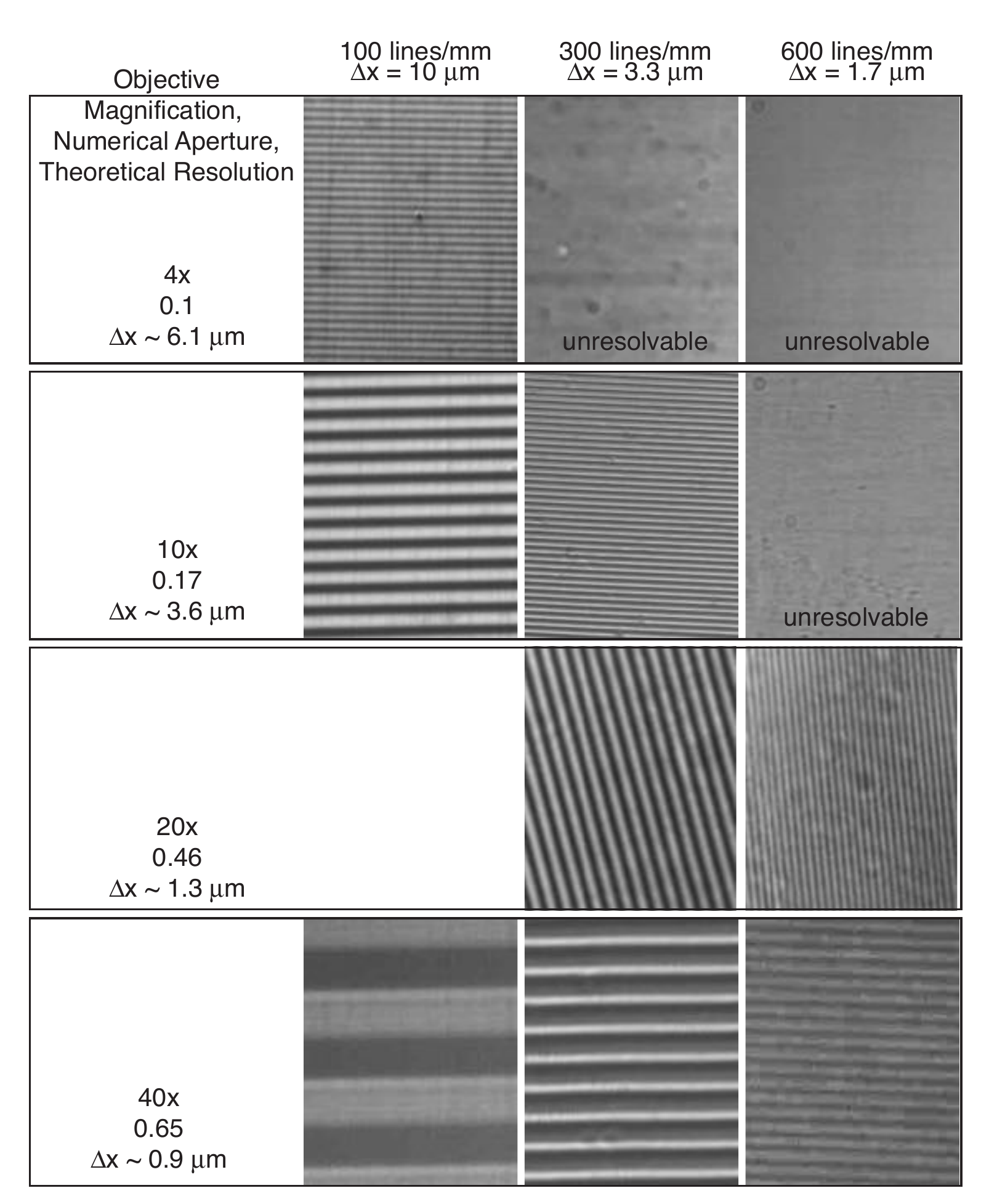}
\caption{Example data taken by students of diffraction gratings with spacings of 100 lines per mm, 300 lines per mm, and 600 lines per mm taken with objectives with magnifications of 4x, 10x, 20x, and 40x. The magnification, numerical aperture, and theoretical minimum spacing able to be determined by each objective is given in the first column. The 4x objective is able to resolve the 100 lines per mm because the distance between lines is 10 $\mu$m, but not the 300 lines per mm or 600 lines per mm with spacings of 3.3  $\mu$m and 1.7 $\mu$m. The 10x objective can resolve the 100 lines per mm, and just barely resolve the 300 lines per mm, but not the 600 lines per mm. The 20x and 40x can resolve the lines from all diffraction gratings.}
\label{fig:resolution}
\end{figure}

After creating an imaging path with the correct magnification, we instruct the students to test the resolution limits of their microscope. We provide several diffraction gratings with decreasing line spacings including 100 lines per mm, 300 lines per mm, and 600 lines per mm. For the long-format optics course, we insist that students image the three diffraction gratings using two different magnification objectives. Typically, lower magnification objectives have lower numerical apertures and thus lower resolution (Fig.~\ref{fig:resolution}). The lines of finer gratings cannot be distinguished by low-power, low numerical aperture objectives. Students learn from texts or via online resources how to determine the theoretical resolution of their objectives using the following equation:
\begin{equation}
d_{min} = 1.22 \frac{\lambda}{NA}.
\end{equation}
where $d_{min}$ is the minimum distance that is resolvable, $\lambda$ is the wavelength of light (500 nm), and $NA$ is the numerical aperture of the objective, see equation (1),  \cite{Pedrotti}. Using the gratings, the students must calculate the spacing between lines and determine if they expect the objective to be able to resolve the lines. For instance, a 4x objective with a numerical aperture of 0.15 should be able to resolve gratings with spacing of  4 $\mu$m or larger. The grating with 100 lines per mm have a spacing of 10 $\mu$m and the grating with 300 lines per mm have a spacing of 3.3 $\mu$m. Thus, we would expect for the 100 lines per mm grating to be resolvable, but not the 300 lines per mm nor the 600 lines per mm (Fig.~\ref{fig:resolution}).

As is often true, the experiment rarely perfectly matches the theory. The students find that, when the theoretical resolution is near the spacing of the grating, they may or may not be able to resolve the spacing (Fig.~\ref{fig:resolution}). Often it depends on the numerical aperture of their condenser and the actual wavelength of light the students are using. We suggest for the students to match the numerical aperture of their condenser to the numerical aperture of the objective to obtain the highest resolution. Further, if the students make the condenser aperture too large or too small compared to the objective, the image becomes too dim or too washed out to resolve. Such example allow us to bring up advanced topics and discuss the microscope system as a whole.

We encourage students to use different objectives with different numerical apertures to test the resolution limits. This allows students to delve deeper into the concept of measurement and how accurately they can measure distance with a light microscope. They can also speculate on other parts that could be altering the resolution of the system they built, such as their condenser or the intensity of light propagating through the system.  

\section{Advanced Topics}
There are several advanced projects or topics that can be done in order to introduce more advanced imaging and analysis techniques. In this section, we describe one additional building project, the epi-fluorescence path, and one analysis technique, super-resolution fitting to gain accuracy in their distance measurements.

\subsection{Epi-Fluorescence Path}
In the optics course where the microscope building was a semester-long project, the final part of the course is for students to build an advanced optical system onto the basic microscope they built. One particularly popular system to build was the epi-fluorescence path. Epi-fluorescence is a heavily-used modern technique in the life sciences. Most modern microscopes have an epi-fluorescence imaging path (Fig.~\ref{fig:modernmicroscope}). Epi-fluorescence imaging is achieved through ``epi'' illumination where the excitation light is shined onto the sample from the same direction that the viewer will image the sample. In other words, the epi-illumination path must pass through the objective. This can be observed in the schematic in figure~\ref{fig:modernmicroscope}, where the epi-fluorescence lamp and dichroic are under the objective.  The reason for this configuration is that the majority of the background photons from the illumination will pass up and away from the imaging side, which will reduce the background fluorescence significantly. 

Additional equipment for epi-fluorescence includes an additional LED illumination source, color filters, and additional lenses and irises, and some method of blocking extra light from the system. The LED source should be of the correct wavelength for fluorescence. We use a green LED for red fluorescence. If a white-light LED is used, one needs to be sure that the desired excitation wavelength is part of the LED's spectrum. Further, you will require a color filter to block the undesired wavelengths in front of the LED. The wavelengths of the filters for excitation that you need depend on the fluorophore you are trying to image. There are a number of resources online to help you find the best filters each fluorophore. \cite{curv-o-matic}

\begin{figure}
\centering
\includegraphics[width=3.5 in]{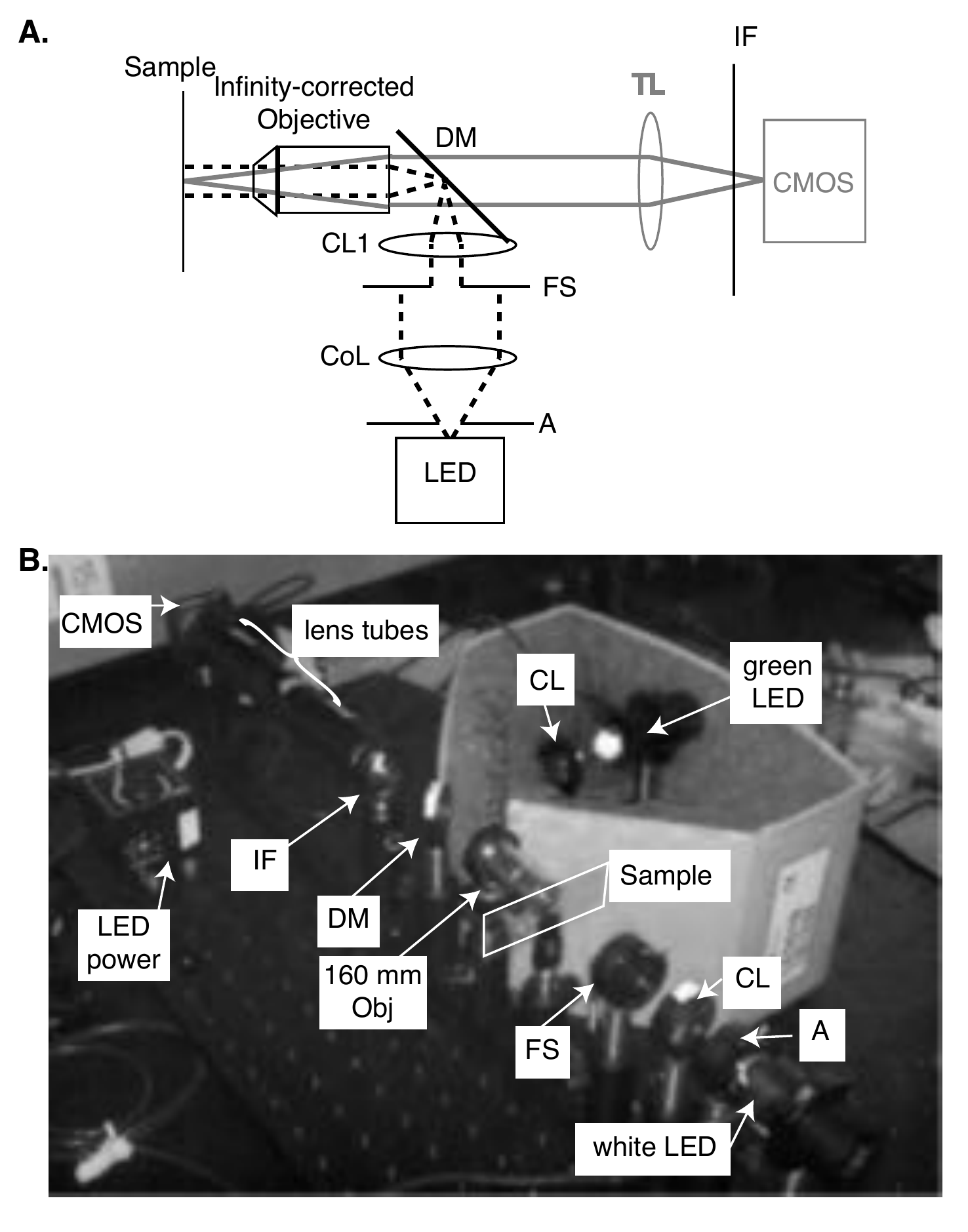}
\caption{Example epi-fluorescence microscope designs created by students. \textbf{A.} Schematic diagram of epi-fluorescence path (the transmitted light condenser is not shown). The diagram represents a three-lens epi-fluorescence condenser that uses the infinity-corrected objective as the third lens. The dashed lines represent the real light rays that are emitted by the LED. LED = Light emitting diode used as the source. We used a green LED, but a white LED can be used with an interference filter to create the correct excitation wavelength. A = Aperture to control the light intensity. CoL = Collecting lens that collects the LED light. FS = Field stop to control the area of illumination. CL1 = Condenser lens 1. DM = Dichroic mirror that reflects excitation light into the sample and allows emission fluorescence wavelengths through to the camera. Solid lines represent the real light rays of fluorescence that come from the sample and are imaged onto the CMOS camera using a tube lens (TL). Before the fluorescence light goes to the camera, we block other wavelengths using an interference filter (IF) that selects the wavelengths of the emission light of fluorescence.  \textbf{B.} Photograph of an epi-fluorescence microscope created by students that uses a 160 mm focal length objective and a single condenser lens (CL) in front of the green LED. They also use a dichroic mirror (DM) that reflects green light into the objective onto the sample. After the dichroic mirror, an interference filter that picks the correct fluorescence wavelength is used before the CMOS camera 160 mm away from the objective. This group also employed a tube lens and cardboard to block stray room light.}
\label{fig:epi}
\end{figure}

In the epi-illumination scheme, you need to reflect the excitation wavelength into the objective and ultimately the sample, but still allow the emitted fluorescence light to go to the camera. In order to achieve the spectral separation, a special dichroic beam-splitter is used, called the dichroic mirror. The dichroic mirror is positioned at a 45 degree angle to reflect the excitation light into the objective (Fig.~\ref{fig:epi}). The optimal dichroic mirror depends on the excitation and emission wavelengths of the fluorophore you choose. \cite{curv-o-matic} 

Once the excitation light hits the sample, the sample will fluoresce, provided the correct fluorophore is present. We recommend using a very bright sample to begin. For green fluorescence using blue excitation light, a cover glass with highlighter marker ink on it will work. Highlighter marker ink has a high concentration of fluorescein, an organic dye that is  bright. For red fluorescence using green excitation light, we use rhodamine dissolved in water or rhodamine-labeled beads. Rhodamine is an inexpensive organic fluorophore that is carcinogenic, so students should be careful handling it. The first sample should have a lot of fluorescent dye, so that the fluorescence can be observed by eye when the excitation light hits the sample. Some of the fluorescent light will shine back through the objective and be directed to the camera.

In order to image the emitted fluorescence light on the camera and block other wavelengths, we need to use an emission filter. The optimal wavelength for emission for your chosen fluorophore can be found. \cite{curv-o-matic} The emission filter should be positioned in front of the camera. Students found that positioning the filter directly in front of the detector, or using black lens tubes helped to reduce stray light. Using the samples with a high density of fluorescent molecules, students should be able to see bright light on the camera even with all the room lights off or the sample completely covered by light-blocking fabric. The fluorescence intensity depends on the amount of excitation light hitting the sample. Students can raise or lower the intensity of the fluorescence LED to make sure that the light on the camera is coming from fluorescence and not stray room light or scattered excitation light. 

Once the students verify that the image on the camera is coming from the high density fluorophore sample, they can perform several tests. For one, the students can alter the excitation and emission filters to determine which are the best for their sample. It is often easy to find inexpensive discontinued or old filters online that filter makers sell on their websites, such as Chroma or Semrock. Others can be found on online auctions.

Another test make a dilute sample with fluorescent beads. We want the number of beads to be relatively low so that individual beads can be visualized. Beads are diluted in water and pipetted into a flow chamber made from a slide, cover glass, and double-stick permanent tape. The chamber can be sealed on the sides using wax or epoxy. Individual fluorescent beads should be visible, but they are much dimmer than the high density sample. If beads are not visible at first, more light blocking may be needed. Students found it best to block stray room lights with cardboard because this contributed to background noise and made it difficult to image. They also tried a variety of dichroic and emission filters to optimize the wavelengths for the beads we supplied. Using the images of fluorescent beads over time, students can perform experiments to measure diffusion and obtain a sense of Boltzmann distribution in a gravitational field, such as those described previously by Peidle and colleagues.\cite{Peidle} 

\subsection{Super-Resolution}
Recently, so-called ``super resolution" techniques have been implemented to improve the imaging capabilities of fluorescence microscopy. Some techniques, such as Stimulated Emission Depletion (STED)\cite{Muller} and Structured Illumination Microscopy (SIM)\cite{Jost} imaging use optics to create patterns of light that are smaller than the diffraction limit. Other methods, such as Photoactivation Localization Microscopy (PALM)\cite{Patterson} or Stochastic Optical Reconstruction Microscopy (STORM)\cite{Huang} image individual fluorescent molecules one at a time and fit each one with high accuracy to find the center of the molecule. Individual molecule locations are overlaid together to create an image, like pointillism.\cite{Scherer} 

The second type of ``super-resolution,'' PALM and STORM, uses a priori knowledge of the intensity profile expected in order to obtain a higher accuracy of localization of individual fluorophores. \cite{superres} For instance, we know beforehand, that the linear optical system of the microscope, takes point-like sources and convolves them with the point-spread function of the objective to create a 2D Bessel function. \cite{Hecht} The Bessel function appears strikingly like a 2D. Gaussian function with the majority of the intensity coming from the center and decaying exponentially at the edges. \cite{Gelles} By fitting the intensity profile of the convolved image of a single point source to a 2D Gaussian, the center of the intensity pattern can be localized well, assuming there are enough photons and the pixel size is neither too small or too large. \cite{Webb}  

Similarly, if the students know a priori knowledge about the sample they are imaging, they will be able to determine the distances between objects with better accuracy than allowed by traditional resolution limits. Here, we present a similar analysis method as PALM or STORM analysis can be used to find the spacing between the lines of a diffraction grating with accuracy far higher than the diffraction-limited uncertainty would allow. The a priori knowledge is that the students are imaging a grating with a repeating structure at a constant spacing. Before we describe the technique, it should be noted that this technique was developed by undergraduate students in our advanced physics laboratory. 

\begin{figure}[h!]
\centering
\includegraphics[width=3.5 in]{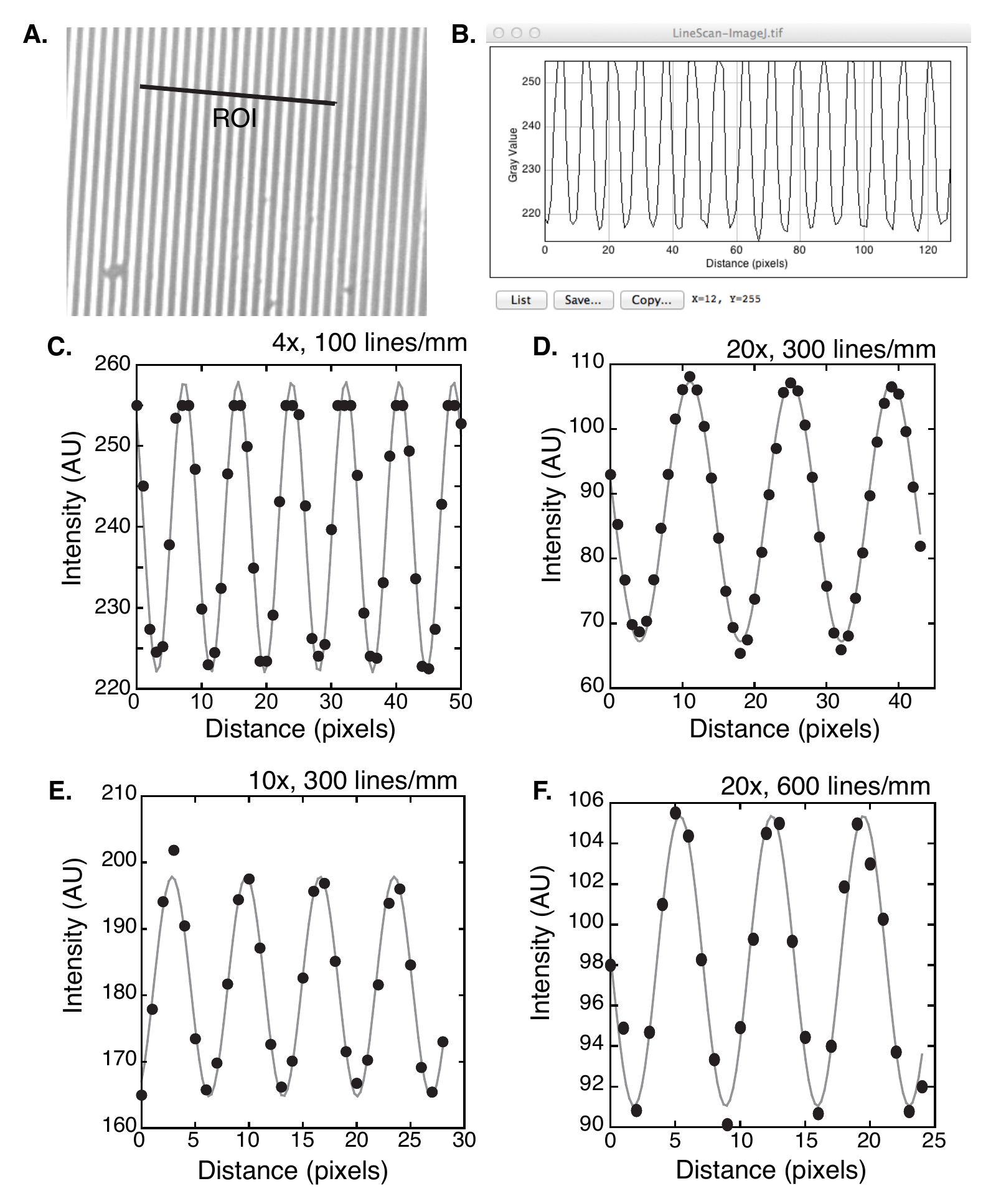}
\caption{Super-resolution fitting of diffraction grating image to achieve accuracy better than the resolution limit. \textbf{A.} An image of a diffraction grating with a line region of interest drawn perpendicular to the grating lines. \textbf{B.} A screenshot from ImageJ showing the linescan of the intensity along the linear region of interest. 
\textbf{C.} Example intensity as a function of distance with given sinusoidal function to the 100 lines per mm grating imaged with the 4x objective. Fit parameters are: $I_0=240.0 \pm 0.1$, $A=17.9 \pm 0.3$, $x_0=-3.07 \pm 0.04$, $\lambda = 2.076 \pm 0.002$,
and the goodness of fit was $R^2$=0.98.
\textbf{D.} Example intensity as a function of distance with sine wave fit to the 300 lines per mm grating imaged with the 20x objective. Fit parameters are: $I_0=87.2 \pm 0.2$, $A=-20.1 \pm 0.3$, $x_0=-13.474 \pm 0.08$, $\lambda = 3.505 \pm 0.008$,
and the goodness of fit was $R^2$=0.99.
\textbf{E.} Example intensity as a function of distance with sine wave fit to the 300 lines per mm grating imaged with the 10x objective. Fit parameters are: $I_0=181.3 \pm 0.3$, $A=16.5 \pm 0.4$, $x_0=1.14 \pm 0.05$, $\lambda = 1.72 \pm 0.06$,
and the goodness of fit was $R^2$=0.99.
\textbf{F.} Example intensity as a function of distance with sine wave fit to the 600 lines per mm grating imaged with the 20x objective. Fit parameters are: $I_0=98.2 \pm 0.2$, $A=-7.12 \pm 0.3$, $x_0=0.06 \pm 0.07$, $\lambda = 1.77 \pm 0.01$,
and the goodness of fit was $R^2$=0.97.}
\label{fig:superres}
\end{figure}

The method to find the spacing at high resolution starts by using ImageJ to open the image of the grating, as used to find the magnification. Instead of measuring the spacing between the lines by drawing a line by hand, we will use the intensity profile across the grating (Fig.~\ref{fig:superres}, A). The intensity profile can be found by drawing a line across several of the grating lines, and then using the Plot Profile tool in the Analyze menu. The plot profile function will create a new window of a plot that shows the intensity as a function of the distance along the line that was drawn (Fig.~\ref{fig:superres}, B). Performing these functions demonstrate to the students the concept that the grayscale observed on the image corresponds to a numerical value. If the image is an 8-bit image, absolute black is a numerical value of 0 and absolute white has a numerical value of 255.

In order to save the numerical array of intensity as a function of distance, the student will need to List the values and save the data as a text file. This file can be opened in Origin, KaleidaGraph, MatLab, Python, or similar programs to be plotted and fit with a sine wave (Fig.~\ref{fig:superres}, C-F). We fit the data to a sine wave of the form:
\begin{equation}
I(x) = I_0 + A \sin(\pi \frac{x-x_0}{2\lambda}).
\end{equation}
where $I(x)$ is the intensity as a function of the distance $x$, $I_0$ is the background intensity, $A$ is the amplitude, $\lambda$ is the wavelength, and $x_0$ is the phase offset in the x-direction.   

The distance between the grating lines is equal to the measured $\lambda$ from the fit. The error of this fit is much smaller than the resolution limit, and thus the measurement is more accurate. The high accuracy is thanks to using extra knowledge of the repeating pattern to fit multiple lines. The magnification can be determined using this same method, and can be found with higher accuracy. The uncertainty for finding the distance between lines of a grating can be smaller (one to tens of nanometers) than the expected resolution of the system, which is approximately half the wavelength of the light (hundreds of nanometers).

 \section{Summary}
 We have described a laboratory experimental set-up for students to build a working light microscope that duplicates the light path of a modern microscope. This equipment can be used as a hands-on project for learning geometric optics and the principles of measurement and uncertainty. We have employed the equipment and tasks described here to a semester-long optics course, an advanced laboratory course for physics majors, and an interdisciplinary graduate lab module.  The equipment we describe is inexpensive; the cost for a complete system where you have nothing to begin with costs  less than \$3000. Parts can often be purchased used, borrowed, or 3D printed making this system easily-acquired. For instance, many optics labs already have screws and other small optomechanics parts, and many microscopists have older model objectives that they simply do not use anymore. The basic system we describe can be further modified to include more advanced systems, such as an optical tweezer.\cite{Smith, Bechoefer, Appleyard}

 \begin{acknowledgments}
This work was supported by a Cottrell Scholars Award to JLR from Research Corporation for Science Advancement and the Department of Physics, University of Massachusetts, Amherst. JLR is also supported by NSF grants INSPIRE-MCB-1344203 and DMR-1207783. RK was a student in the Optics for Biophysics course BIO577/578 in Spring 2013, AC, JS, AT, SZ were students in the advanced laboratory course for physics majors PHYS440 in Fall 2013. Each of these students contributed data, images, and edited the manuscript. For more information about the course Optics for Biophysics, please email JLR.
\end{acknowledgments}

\begin{table}[h!]\small
     \begin{center}
     \begin{tabular}{ |c |p{2cm} | p{6cm} | p{6cm} | }
     \toprule
     \hline
     
       Image & Part No. & Description & Use  \\
       \hline
    \cmidrule(r){1-1}\cmidrule(lr){2-2}\cmidrule(l){3-3}\cmidrule(r){4-4}
     \raisebox{-\totalheight}{\includegraphics[scale=0.15]{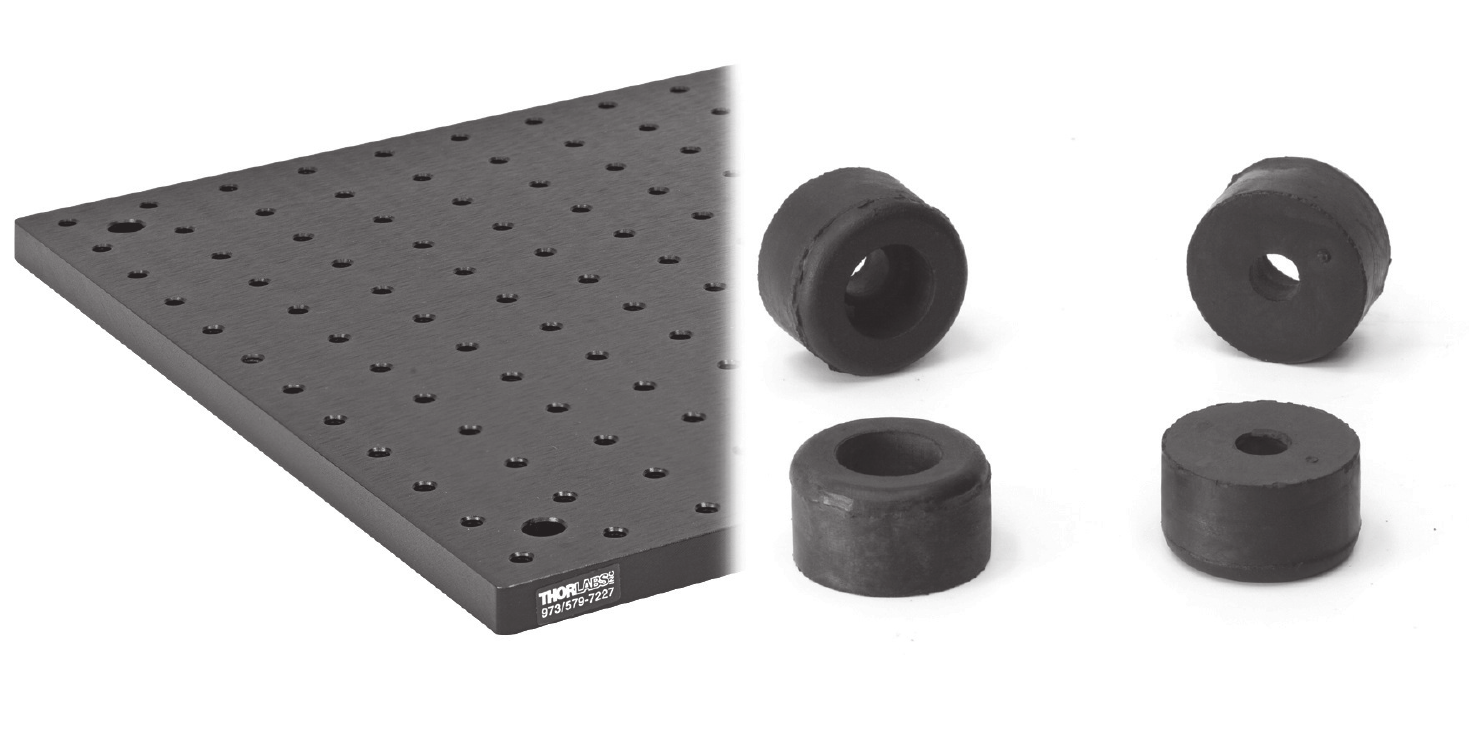} }
      & MB2436, RDF1 &
       Optics Breadboard with rubber feet (2' x 3', 1" tapped 1/4-20 holes) & Building surface and minor vibration isolation \\
     \hline

    \cmidrule(r){1-1}\cmidrule(lr){2-2}\cmidrule(l){3-3}
     \raisebox{-\totalheight}{\includegraphics[scale=0.1]{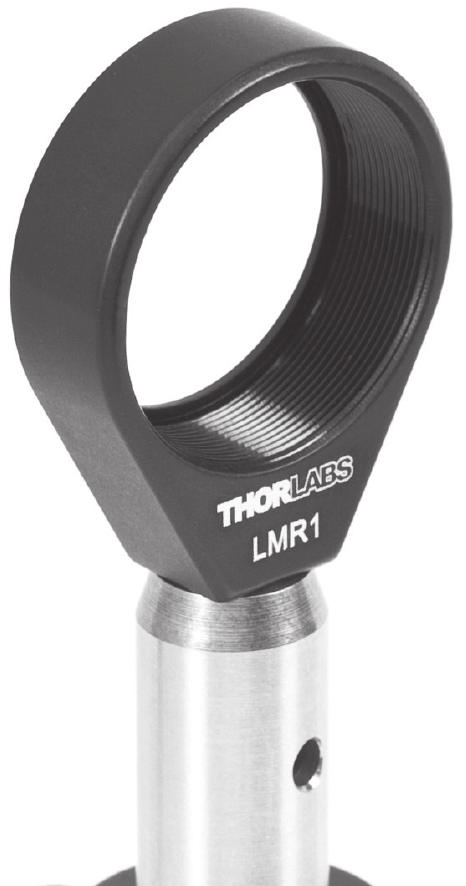} }
      & LMR1 &
     1-inch lens mounts & To hold 1-inch lenses \\
          \hline
           
           \cmidrule(r){1-1}\cmidrule(lr){2-2}\cmidrule(l){3-3}
     \raisebox{-\totalheight}{\includegraphics[scale=0.2]{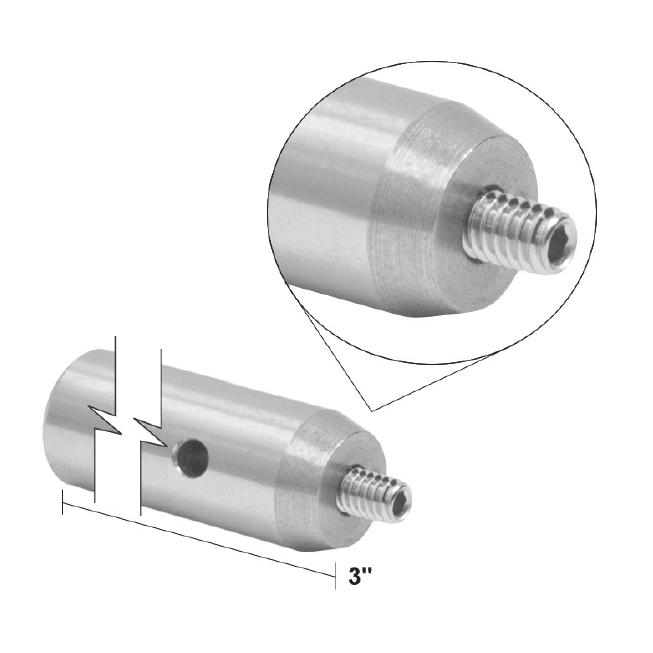} }
      & TR2 &1-inch wide posts (2-inch long) & To mount to lens holders \\
          \hline
          
          \cmidrule(r){1-1}\cmidrule(lr){2-2}\cmidrule(l){3-3}
     \raisebox{-\totalheight}{\includegraphics[scale=0.05]{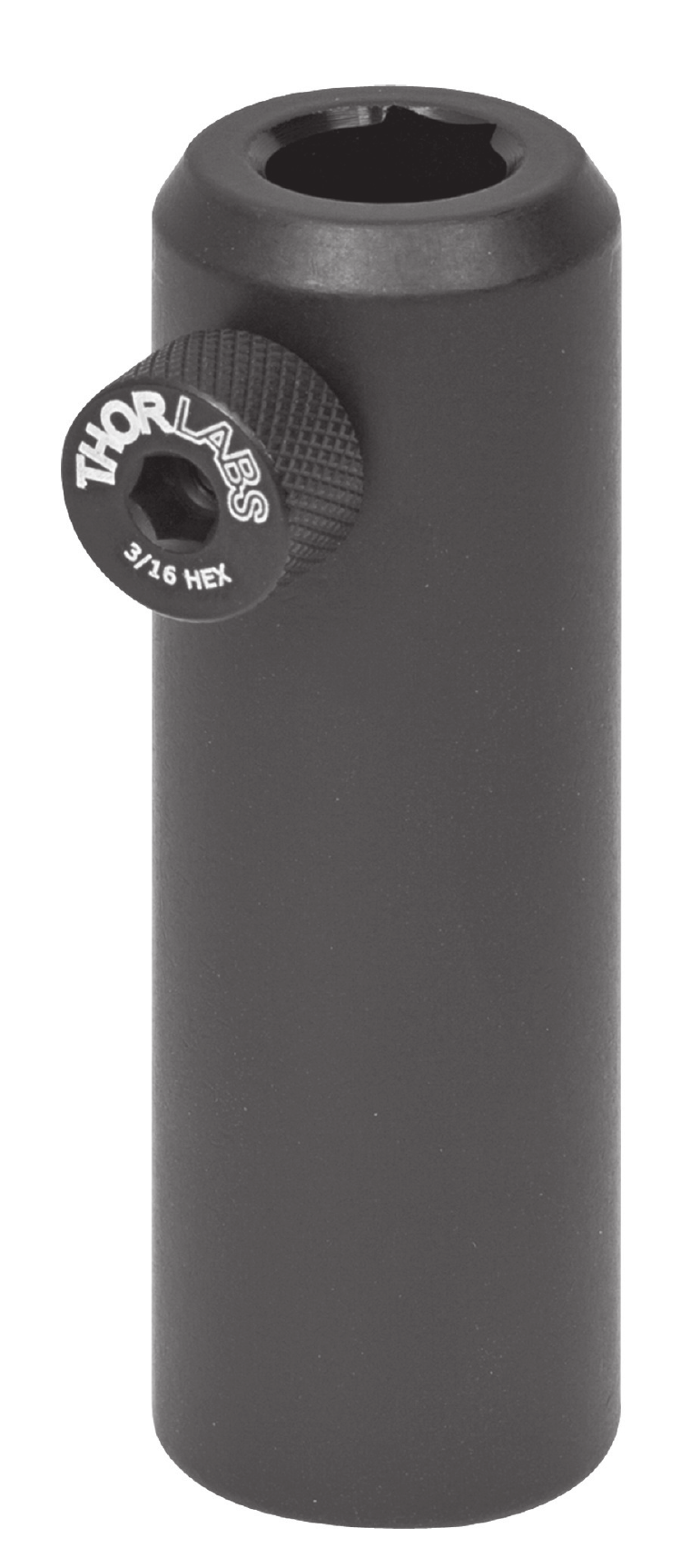} }
      & PH2 &Post-holders (2-inch long)  & To hold 1-inch posts with adjustable height  \\
          \hline
          
          \cmidrule(r){1-1}\cmidrule(lr){2-2}\cmidrule(l){3-3}
     \raisebox{-\totalheight}{\includegraphics[scale=0.15]{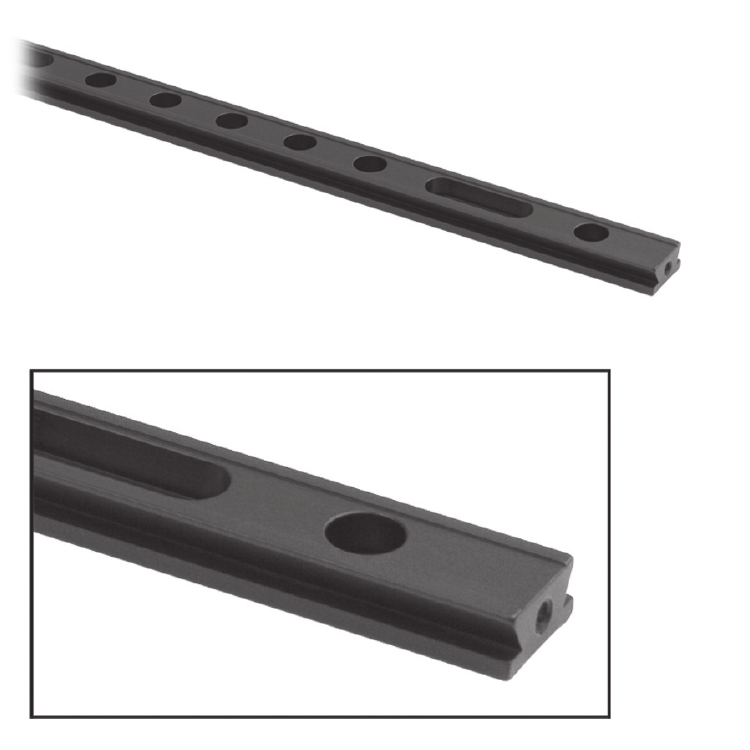} }
      & RLA1200 &Dove tail rail (2 x 12-inches long)  & To mount optics with dove tail carts for adjustment along optical axis \\
          \hline
          
          \cmidrule(r){1-1}\cmidrule(lr){2-2}\cmidrule(l){3-3}
     \raisebox{-\totalheight}{\includegraphics[scale=0.15]{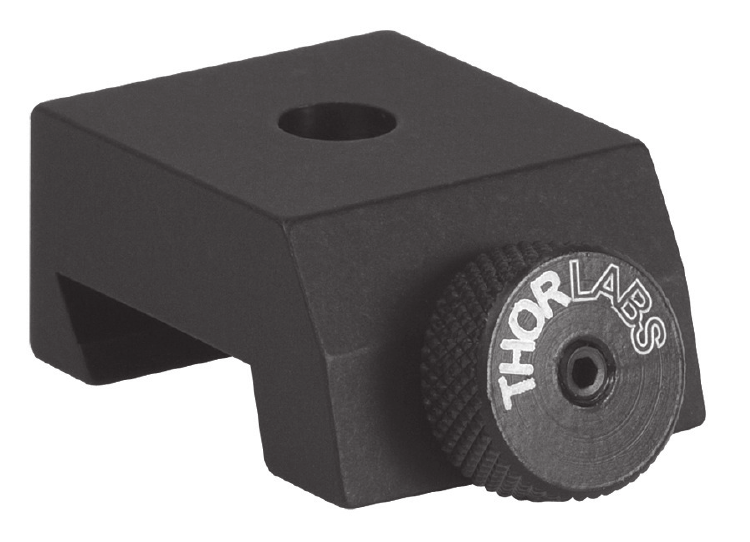} }
      & RC1 &Dove tail rail carts & To mount post holders slide along optical axis \\
          \hline
          
          \cmidrule(r){1-1}\cmidrule(lr){2-2}\cmidrule(l){3-3}
     \raisebox{-\totalheight}{\includegraphics[scale=0.15]{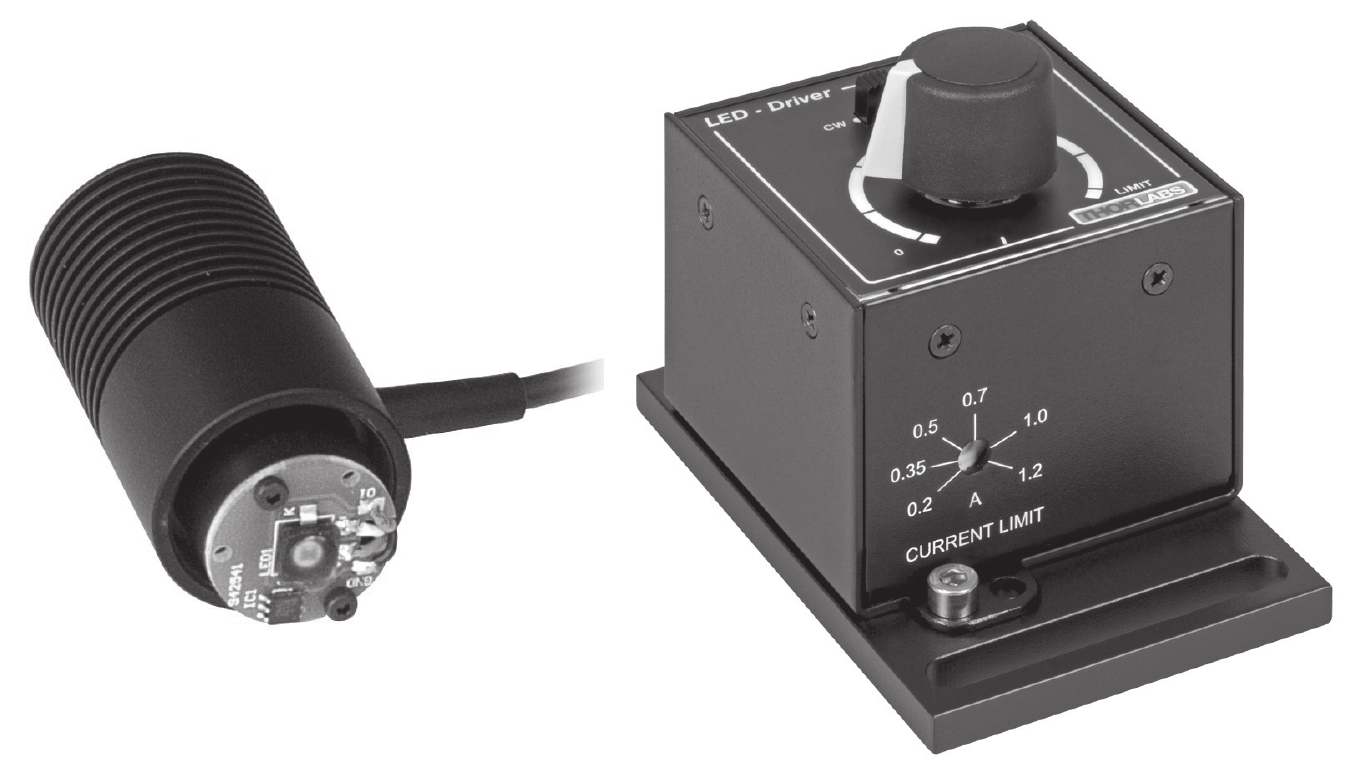} }
      & MCWHL5, LEDD1B &
      White LED with power controller (does not come with power adapter cable) & Transmitted light illumination source \\
          \hline
          
          \cmidrule(r){1-1}\cmidrule(lr){2-2}\cmidrule(l){3-3}
     \raisebox{-\totalheight}{\includegraphics[scale=0.3]{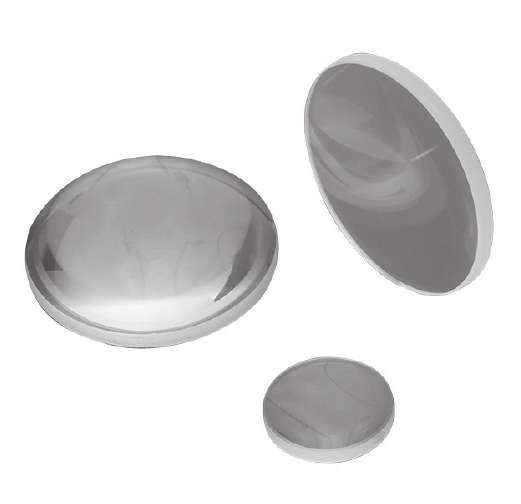} }
      & LA1255-A &Various Lenses (f = 30 mm, 50 mm, 100 mm, 200 mm, plano-convex,  achromatic) & To build condenser system and tube lens for imaging  \\
          \hline
          
          \cmidrule(r){1-1}\cmidrule(lr){2-2}\cmidrule(l){3-3}
     \raisebox{-\totalheight}{\includegraphics[scale=0.1]{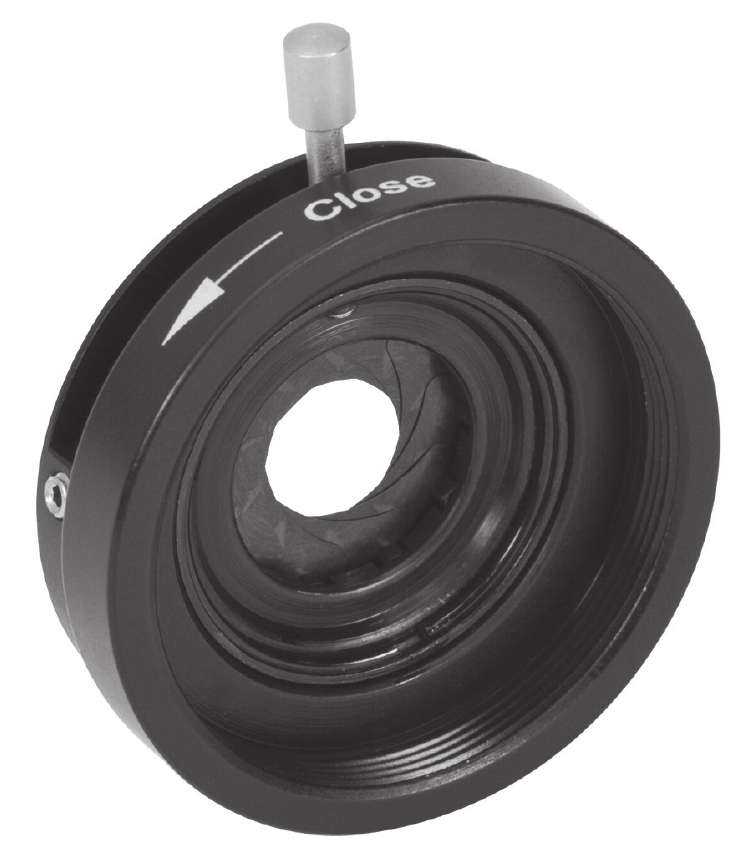} }
      & SM1D12 &Variable Irises & To control light as field stops and apertures  \\
          \hline
          
          \cmidrule(r){1-1}\cmidrule(lr){2-2}\cmidrule(l){3-3}
     \raisebox{-\totalheight}{\includegraphics[scale=0.15]{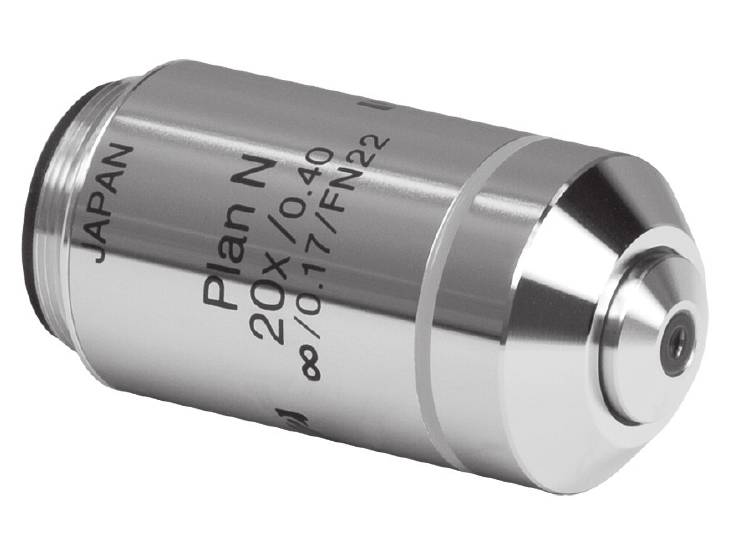} }
      & RMS4X &Objectives (Infinity corrected and 160 mm, 5x, 10x, 20x)  & Imaging system  \\
          \hline
          
          \cmidrule(r){1-1}\cmidrule(lr){2-2}\cmidrule(l){3-3}
     \raisebox{-\totalheight}{\includegraphics[scale=0.15]{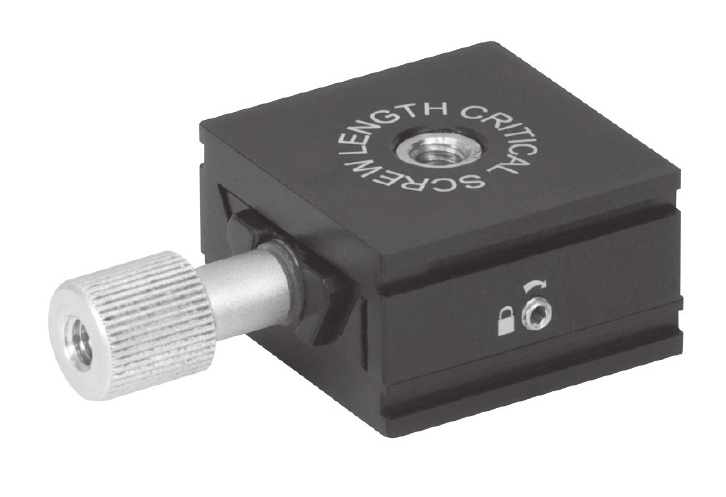} }
      & DT12XY &1-D translation stage & Fine focus control of objective \\
          \hline
          
           \cmidrule(r){1-1}\cmidrule(lr){2-2}\cmidrule(l){3-3}
     \raisebox{-\totalheight}{\includegraphics[scale=0.15]{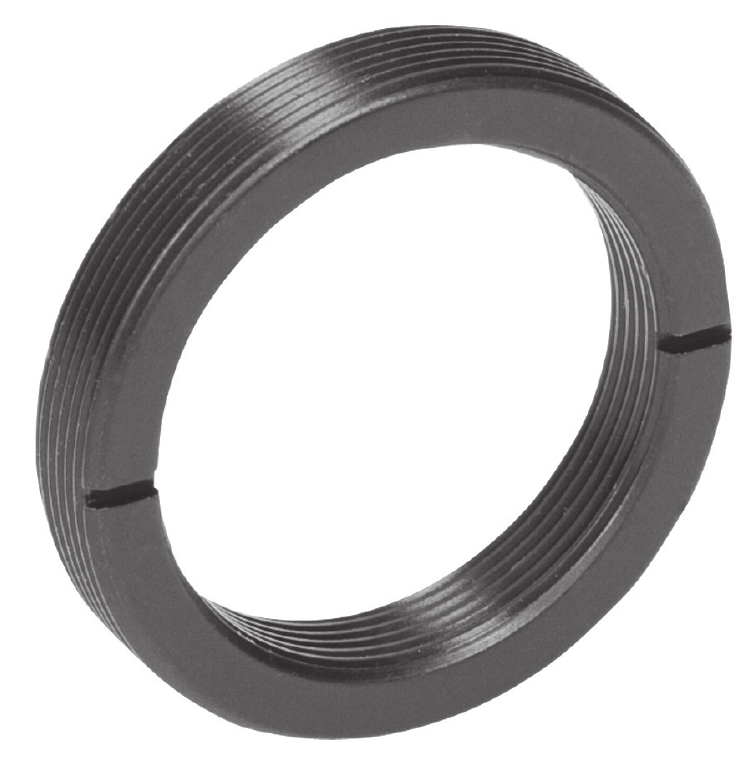} }
      & SM1A3 &1-inch tube lens thread (SM1) to RMS adapter & To hold objective in 1-D translation stage \\
          \hline
          
           \cmidrule(r){1-1}\cmidrule(lr){2-2}\cmidrule(l){3-3}
     \raisebox{-\totalheight}{\includegraphics[scale=0.15]{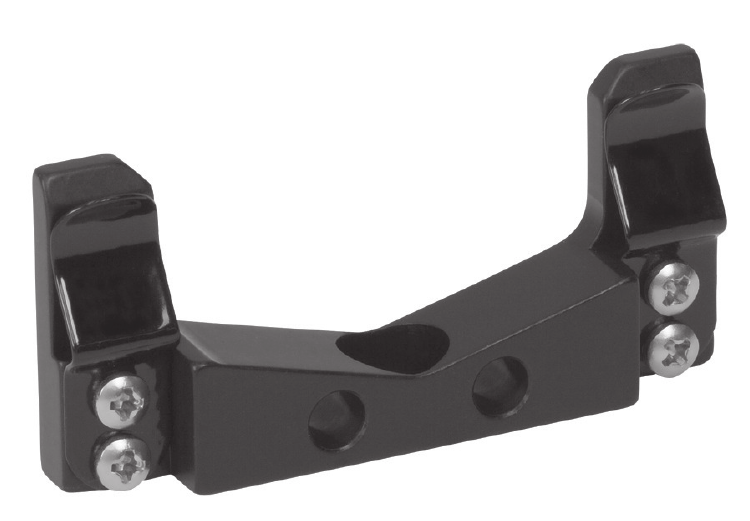} }
      & FH2D &Pressure filter holder & To hold the sample slide \\
          \hline
          
           \cmidrule(r){1-1}\cmidrule(lr){2-2}\cmidrule(l){3-3}
     \raisebox{-\totalheight}{\includegraphics[scale=0.15]{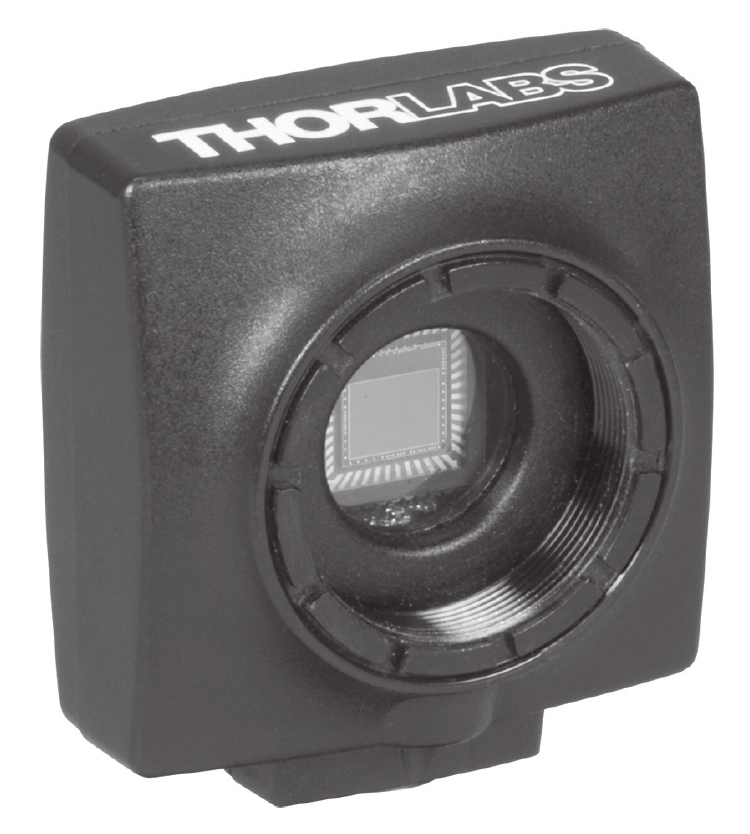} }
      & DCC1545M &CMOS-USB camera & To send electronic image to the computer \\
          \hline
          
           \cmidrule(r){1-1}\cmidrule(lr){2-2}\cmidrule(l){3-3}
     \raisebox{-\totalheight}{\includegraphics[scale=0.15]{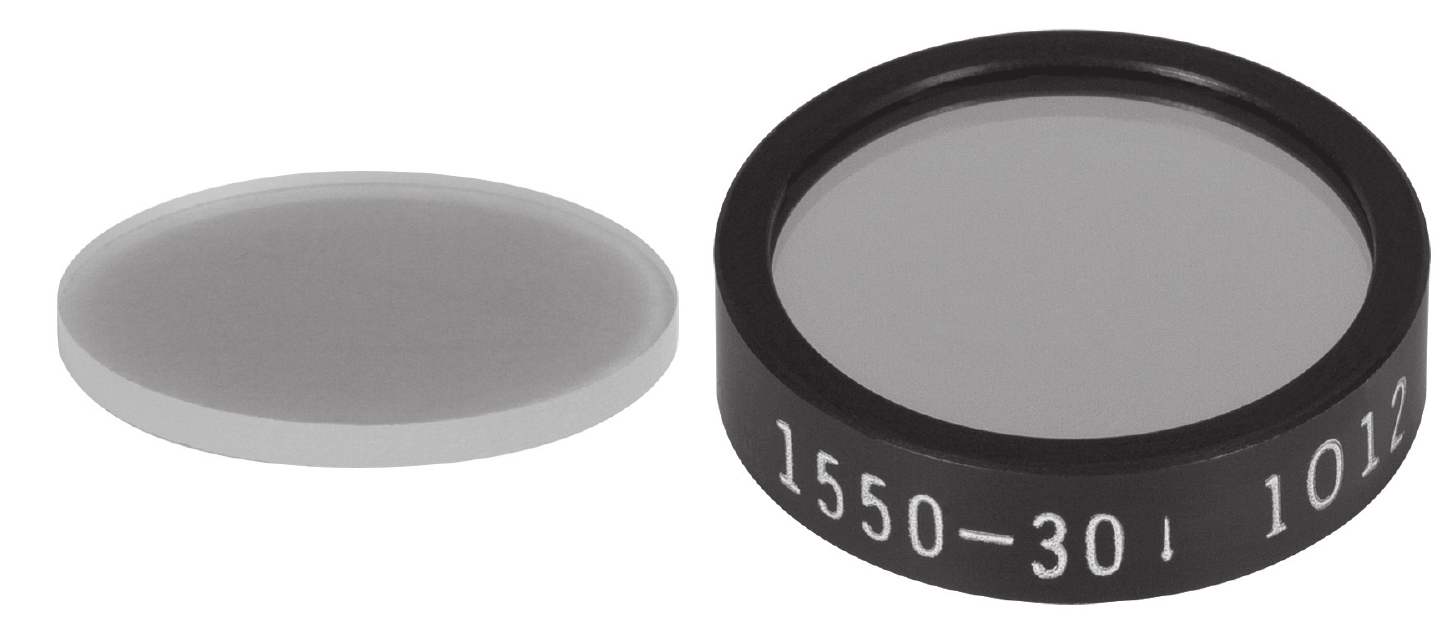} }
      & MF542-20, MD568, MF620-52 &Dichroic filter and colored interference filters & To pick wavelengths for fluorescence. The exact excitation, dichroic, and emission filter depends on the fluorophore.  \\
          \hline
          
       \bottomrule
      \end{tabular}
      \caption{List of optical components with images. Part numbers and images are from ThorLabs, used with permission.}
      \label{Components}
      \end{center}
      \end{table}

\end{document}